\documentclass[aps,prb,twocolumn]{revtex4-2}

\usepackage{amsmath}
\usepackage{amssymb}
\usepackage{graphicx}
\usepackage{epstopdf}
\usepackage[colorlinks=true]{hyperref}
\usepackage{xcolor}
\usepackage{scalerel}
\usepackage[utf8]{inputenc}
\usepackage{amsfonts}
\usepackage{graphicx}
\usepackage{dcolumn}
\usepackage{bm}
\usepackage{hyperref}
\usepackage{accents}
\usepackage{ gensymb }

\usepackage{mathtools}

\newcommand{\bd}[1]{\mathbf{#1}}
\newcommand{\pmat}[1]{\begin{pmatrix} #1 \end{pmatrix}}

\newcommand{\ld}{\lambda_{\mathrm{d}}}
\newcommand{\ls}{\lambda_{\mathrm{s}}}
\newcommand{\ms}{M_{\mathrm{s}}}

\newcommand{\A}{\frac{\Theta_{11}}{\mathcal{M}_{11}} }
\newcommand{\B}{\frac{\Theta_{22}}{\mathcal{M}_{22}} }

\usepackage{times}
\usepackage{setspace}

\begin{document}
\title{Current-driven dynamics of antiferromagnetic domain-wall skyrmions}

\author{Wooyon Kim}
\affiliation{Department of Physics, Korea Advanced Institute of Science and Technology, Daejeon 34141, Republic of Korea}
\author{Jun Seok Seo}
\affiliation{Department of Physics, Korea Advanced Institute of Science and Technology, Daejeon 34141, Republic of Korea}
\author{Se Kwon Kim}
\email{sekwonkim@kaist.ac.kr}
\affiliation{Department of Physics, Korea Advanced Institute of Science and Technology, Daejeon 34141, Republic of Korea}

\begin{abstract}
Domain-wall skyrmions are magnetic solitons embedded in a domain wall that are topologically equivalent to skyrmions. Here, we theoretically study antiferromagnetic domain-wall skyrmions and their current-driven motion within the Landau-Lifshitz-Gilbert phenomenology, and verify our findings with micromagnetic simulations. While the skyrmion Hall effect is expected to be suppressed in the current-induced motion of antiferromagnetic domain-wall skyrmions, we observe a finite Hall angle, which originates from the anisotropic spin configuration of domain-wall skyrmions. The skyrmion Hall effect is, however, conditionally suppressed and the motion aligns with the current applied in certain directions, which can be interpreted as principal axes of a domain-wall skyrmion that is easily identified from the symmetry of the spin configuration. Our work on antiferromagnetic domain-wall skyrmions shows that the dynamics of spin textures endowed with multiple soliton characteristics can be unconventional, which is envisaged to enrich the field of topological solitons.
\end{abstract}
\maketitle

\section{Introduction} \label{Sec:Intro}

The richness of topological solitons in magnetic materials has drawn great attention from researchers with both fundamental and practical interests~\cite{KosevichPR1990, GalkinaLTP2018}. The wide variety of solitons in different classes of magnets offers a vast resource for applied scientists, while the succinct explanation of their dynamics prescribed by the underlying topology evokes an instinctive interest of theorists. The current-driven motion of solitons is not only a topic for spintronic device applications but also a fundamental study of the interaction between charge and spin textures. In this way, the discovery of a new soliton can have significance in both theoretical and practical aspects. One example may include a vertical Bloch line, which was found to evolve inside a domain wall (DW) to suppress Walker breakdown~\cite{Kab}. On the other hand, it was proposed as a building block enabling a memory device beyond the renowned DW-based racetrack memory~\cite{racetrack}, featuring its fast motion without edge roughness by using a DW as a data-bit channel~\cite{yang}. Indeed, finding and investigating new solitons remains a largely open area to those who are willing to see the potential in them.

The magnetic object of our interest in this work is a soliton called a DW skyrmion. It is a DW substructure that can move inside a DW, featuring versatility in that only an easy-axis anisotropy and Dzyaloshinskii-Moriya interaction (DMI) are required for its viability. Previous research has focused on the ferromagnetic (FM) case~\cite{Cheng, Nitta, Je}, leaving the AFM case unexplored despite its value described as follows. First, DW skyrmions would serve as an ideal starting point if one is to realize DW-substructure-based memory in AFMs, since changing the platform from FM to AFM leaves few DW substructures available due to the absence of demagnetization in AFMs. For example, the aforementioned vertical Bloch lines are not supported in AFMs, while they can be stabilized in FMs by demagnetization. Second, as will be shown below, AFM DW skyrmions exhibit a unique skyrmion Hall effect, unlike the typical Hall effect observed in FMs. Due to the compensated topological charge, AFM solitons are not predestined to exhibit the skyrmion Hall effect, if any, the effect is not based on the topological property. AFM DW skyrmions provide a good example that shows the topology-independent skyrmion Hall effect in AFMs and its potential practical advantage. 

In this paper, we study a DW skyrmion and its current-driven motion in AFMs within the Landau-Lifshitz-Gilbert framework~\cite{LLG}, and confirm the result through micromagnetic simulation~\cite{mumax3(1),mumax3(2)}.  In Sec.~\ref{Sec:model}, we describe our model system and further investigate the structure of an AFM DW skyrmion. In Sec.~\ref{Sec:Dynamics}, we study its motion induced by electric current, where both spin-transfer torque (STT)~\cite{stt1,stt2,stt3,stt4} and spin-orbit torque (SOT)~\cite{sot1,sot2,sot3,sot4,sot4,sot5,sot6,sot7} are considered. There, we specify the distinctive features of the skyrmion Hall effect observed in the AFM DW skyrmion. In Sec.~\ref{sec:last}, we conclude our work.

\section{Model}\label{Sec:model}
In this section, we study the structure of an AFM DW skyrmion, which we previously introduced simply as a DW substructure. Our model system is a quasi-two-dimensional collinear AFM film with easy-axis anisotropy along the film normal (set as $\hat{\bd z}$). We employed the staggered magnetization $\bd n (\bd r ) = (\bd m _1 (\bd r) - \bd m_2 (\bd r ))/2$ as the main order parameter that is supplemented by the net magnetization $ \bd m (\bd r) = (\bd m_1 (\bd r) + \bd m_2 (\bd r))/2$, where $\bd m_1 (\bd r)$ and $\bd m_2 (\bd r)$ are sublattice magnetization vectors of unit lengths. We consider the energy given by~\cite{Cheng, Dasgupta}
\begin{widetext}
\begin{equation}
U \left[ \bd n ( \bd r ), \, \bd m (\bd r) \right] = \Delta z \iint \left\{ A ( \nabla \bd n )^{ \, 2} - K n_z ^2 + D [ \, n_z ( \nabla \cdot \bd n ) - ( \bd n \cdot \nabla ) n_z \, ]+ \frac{ | \bd m | ^2 }{2 \chi}  \right\} \, \mathrm{d}^2 \bd r \, . \label{Eq:energy}
\end{equation}
\end{widetext}
Here, $\Delta z$ is the film thickness; the exchange interaction, easy-axis anisotropy, and interfacial DMI are considered, with constants $ A > 0, \, K > 0 ,$ and $D$ (set positive, $D > 0 \, ,$ throughout the paper), respectively. The last term with $\chi$, the paramagnetic susceptibility, represents an energy penalty for having finite magnetization $\bd m$~\cite{Dasgupta}. Note the versatility of the model: Only $K$- and $D$-terms are added onto the essential terms $A$ and $\chi$.

\begin{figure}
    \centering
    \includegraphics[width=0.6\columnwidth]{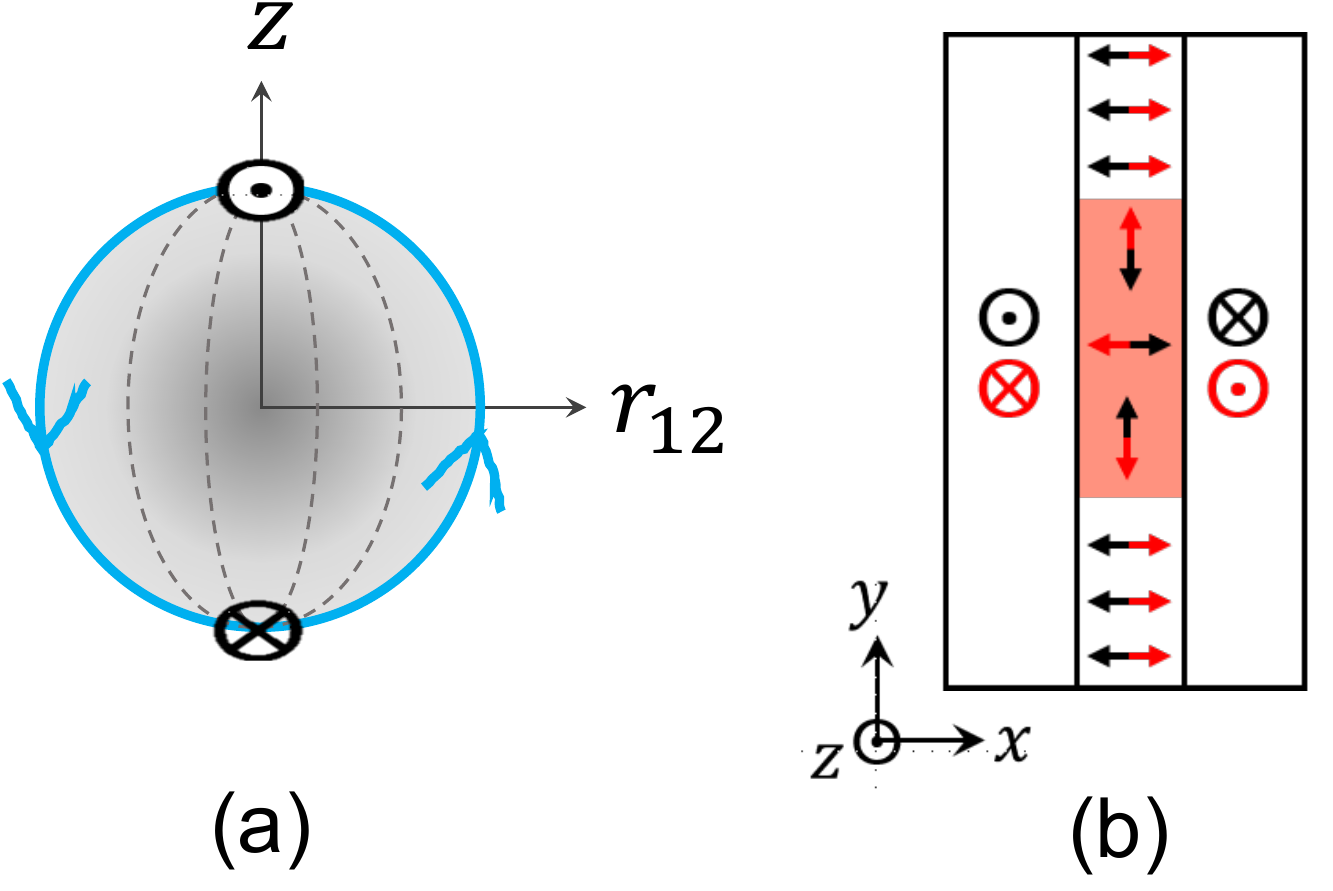}
    \caption{(a) Illustration of a DW that is energetically favored by DMI in order-parameter space. A unit sphere in the order-parameter space, $\bd n \in
    S^2$, is shown overlaid by real-space axes, film normal $\hat{\bd z}$ and separation vector $\bd r_{12}$ between the two positions $\mathbf{r}_1$ and $\mathbf{r}_2$.  The two ground states ($\bd n = \pm \hat{\bd z}$) that are located separately by $\bd r_{12}$ in real space prefer to be interpolated by the blue directed meridians, i.e.,  the counterclockwise rotation about an axis of ``$\, \mathrm{sgn}(D) \, \bd r_{12} \times \hat{\bd z} \, ,"$ where $D$ is DMI coefficient in Eq.~\eqref{Eq:energy}. Other longitudinal lines (dotted) represent less favored DW by DMI. (b) Schematic of an antiferromagnetic DW skyrmion: top view along the film normal $\hat{\bd z}$. The antiparallel black and red arrows represent the magnetizations $\mathbf{m}_1$ and $\mathbf{m}_2$ of the two sublattices. The order parameter $\mathbf{n}$ can be taken to be the equivalent to the black one.
   }
   \label{fig:dwsk-diagram}
\end{figure}

\begin{figure}
    \centering
    \includegraphics[width=\columnwidth]{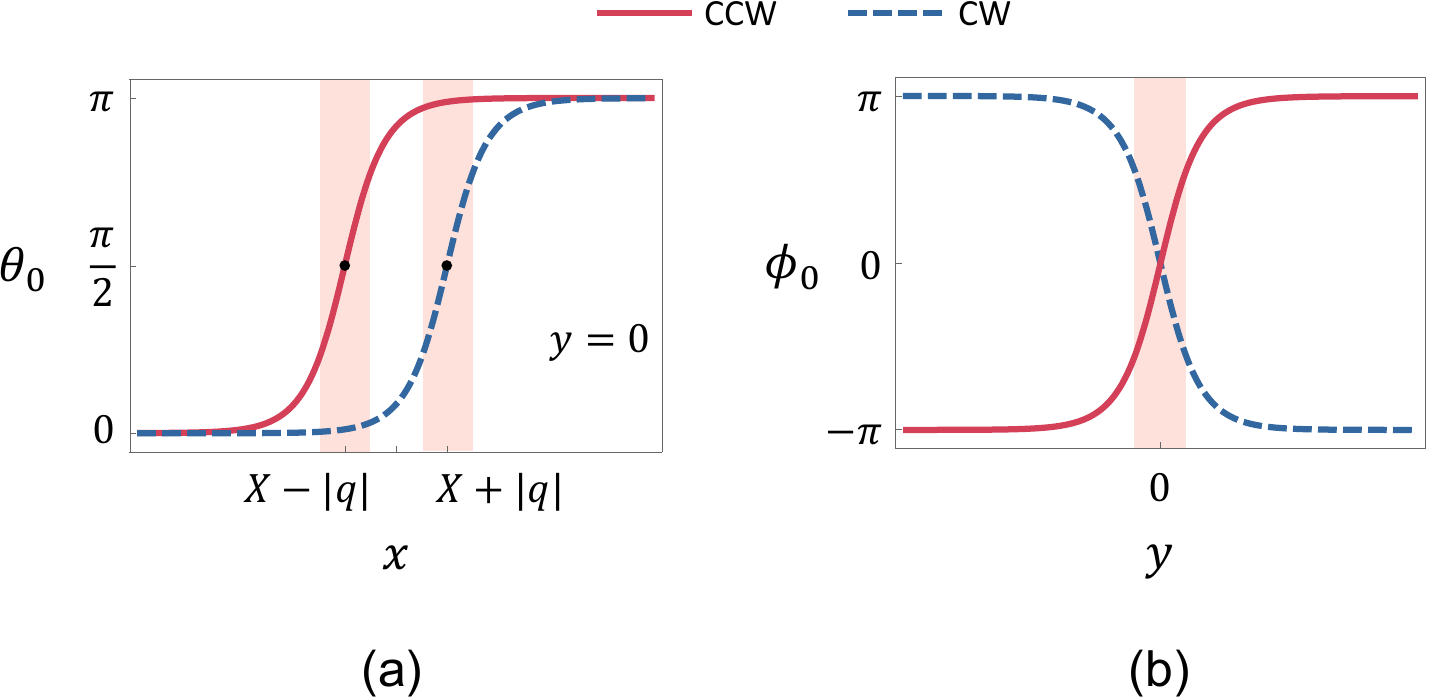}
    \caption{Profile functions of an antiferromagnetic DW skyrmion [Eq.~\eqref{Eq:profile}] with two different types of $\phi_0(y)$ winding: counter-clockwise winding (CCW) with $\phi'_0 (y) > 0$ and clockwise winding (CW) with $\phi'_0 (y) < 0$. (a) Polar angle $\theta_0 (x, y)$ across the midline of a DW, $y = 0 $, and (b) azimuthal angle $\phi_0(y)$. For the two types of DW skyrmions (solid and dotted lines), the DW center shifts in opposite directions, $\mp |q_0|$, and the winding is in opposite directions, $\mathrm{sgn}(\phi_0' )= \pm 1$. The colored regions in the plot indicate the areas within (a) the DW width $\ld$ and (b) the skyrmion radius $\ls$.}
    \label{fig:profile}
\end{figure}

Under the system energy~\eqref{Eq:energy}, the two ground states, $\bd n = \hat{\bd z} $ and $\bd n = - \hat{\bd z} \, ,$ tend to be interpolated in one way around not the other, which is determined by DMI. In order-parameter space $\bd n \in S^2$, this corresponds to a specific directed meridian as shown in Fig.~\ref{fig:dwsk-diagram}(a). A DW skyrmion, however, embraces every possible yet unfavorable meridian, each at once at the expense of a chiral kink~\cite{kink} and energy $\pi |D|\cdot \Delta z$. Altogether, the whole path wraps around a unit sphere in order-parameter space once, having skyrmion charge $Q_{\bd n} = (1/4\pi) \int dxdy [\mathbf{n} \cdot (\partial_x \mathbf{n} \times \partial_y \mathbf{n})]$ of either $1$ or $-1$ [Fig.~\ref{fig:profile}(a)]. The schematic of a DW skyrmion in real space is shown in Fig.~\ref{fig:dwsk-diagram}(b). See closely how the entire meridian in order-parameter space is mapped onto real space: Each meridian on the sphere stretches along the $x$-axis in the plane and is progressively stacked in the $y$-direction, resulting in a $2\pi$-winding inside a DW. A quantitative description of the configuration can be obtained by the energy [Eq.~\eqref{Eq:energy}] minimization with Slonczewski ansatz~\cite{Slonczewski1,Slonczewski2,Cheng} subject to a boundary condition of the preferred DW. In angular representation where $\bd n  = (\sin \theta \cos \phi, \, \sin \theta \sin \phi, \, \cos \theta ) \, ,$ a DW skyrmion solution $\bd n_0 (\bd r)$ is given by
\begin{subequations}
\label{n0}
\begin{align}
\theta_0 ( x, \,  y )
&= 2 \arctan \left[ \exp \frac{ (x - X) - q_0 (y) }{\ld} \right] \, , \label{Eq:theta0}\\
\phi_0 ( y )
&= \pm \Big[ - \pi + 4 \arctan \left( \exp \frac{y - Y }{\ls} \right) \, \Big] \, , \label{Eq:phi0} \\
q_0 ( y )
&= - (\ld) ^2 \frac{\mathrm{d} \phi _0 }{\mathrm{d} y} \, . \label{Eq:q0}
\end{align} \label{Eq:profile}
\end{subequations}
Here, we employed the right-handed Cartesian coordinate system and Euler angles: We set $\hat{\bd x}$ as a DW normal so that the domains of each ground states lie at the ends of the $x$-axis, and $\hat{\bd z}$ as both the polar axis and the film normal; The azimuthal angle is set zero in the $x$ direction.

\begin{figure*}
    \centering
    \includegraphics[width=1.8\columnwidth]{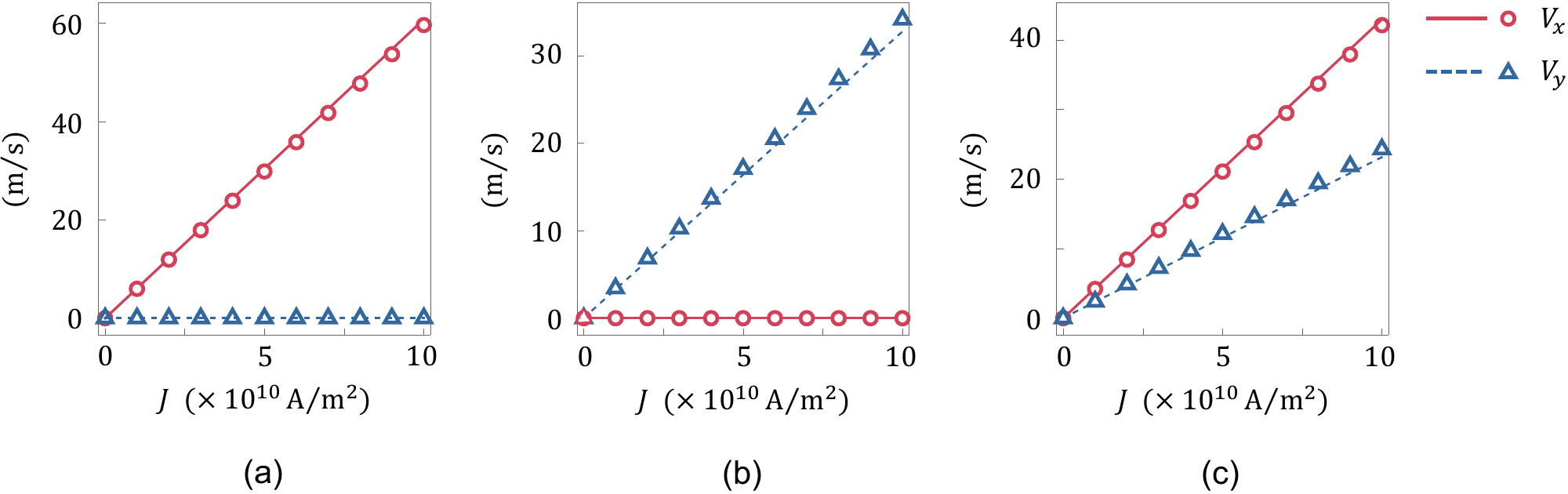}
    \caption{ Switching of the skyrmion Hall effect depending on current direction. Theoretical results (lines) [Eq.~(\ref{Eq:steadyV})] and simulation results (markers) for the motion of a DW skyrmion driven by an electric current. (a)-(c) Velocity components $V_x$ and $V_y$ of a DW skyrmion for a given current density of magnitude $J$. The current is applied (a) along the DW normal (set as $\hat{\bd x}$), (b) along the DW parallel (set as $\hat{\bd y}$), and (c) at an angle $\varphi = 45\degree$ with respect to $x$ axis. The skyrmion Hall effect is suppressed in (a) and (b), while exhibited in (c).
    }
    \label{Fig:skhe}
\end{figure*}

Let us highlight the key features of the profile functions $\theta_0 (x , \, y), \, \phi_0 (y) ,$ and $q_0 (y)$, referring to the plots in Fig.~\ref{fig:profile}. The function $\theta_0$ takes the values 0 and $\pi$ at the far ends, i.e., $|x - X| \gg \ld \, ,$ and monotonically increases in between. Note the translation of a set of DW centers, where $\theta_0 = \pi / 2 : \,\,x = X \rightarrow X + q_0 (y ) \, ,$ which is a manifestation of a chiral kink~\cite{Slonczewski1,Slonczewski2,Cheng}. The function $\phi_0$ takes the value $-\pi$ at the far ends, $ | y - Y | \gg \ls$, hosting the DW that is energetically favored by DMI, while the less-favored DWs by DMI are realized in between. Note the double sign, suggesting that there are two types of a DW skyrmion: One changes its $\phi$-value from $-\pi$ to $\pi$ (as $y$ increases), a counter-clockwise winding, the other changes from $\pi$ to $-\pi$, a clockwise winding. This variation comes from the sequence of implementing the meridians in the order-parameter space into the real system, thus having the same energy cost and both are allowed. The length dimensional function $q_0$, which describes a chiral kink, is accompanied by finite $\mathrm{d} \phi_0 / \mathrm{d} y \, ,$ that is, being away from staying in the most favored DW, $\phi_0 = - \pi$. This supports that the internal skyrmion structure arises at the expense of the kink. Note the direction of a kink and of a winding is correlated as well: $\mathrm{sgn} ( q_0 ) = - \mathrm{sgn} ( \partial_y \phi_0 )$. Lastly, we want to point out the three characteristic elements on geometry: Center of a DW skyrmion $\bd R = ( X, \, Y )$, an intersection of $\theta ( x , \, y ) = 0$ and $\phi (y) = 0 \,$; the DW width $\ld = \sqrt{ A / K} $, which is invariant under the presence of DMI; the DW skyrmion radius $\ls = \ld \cdot \big( \sqrt{ A K } / (\pi D )  \big) ^ { 1 / 2} $, the range along the $y$-axis within which the skyrmion is extended.

\section{Dynamics} \label{Sec:Dynamics}

\begin{figure}
    \centering
    \includegraphics[width=\columnwidth]{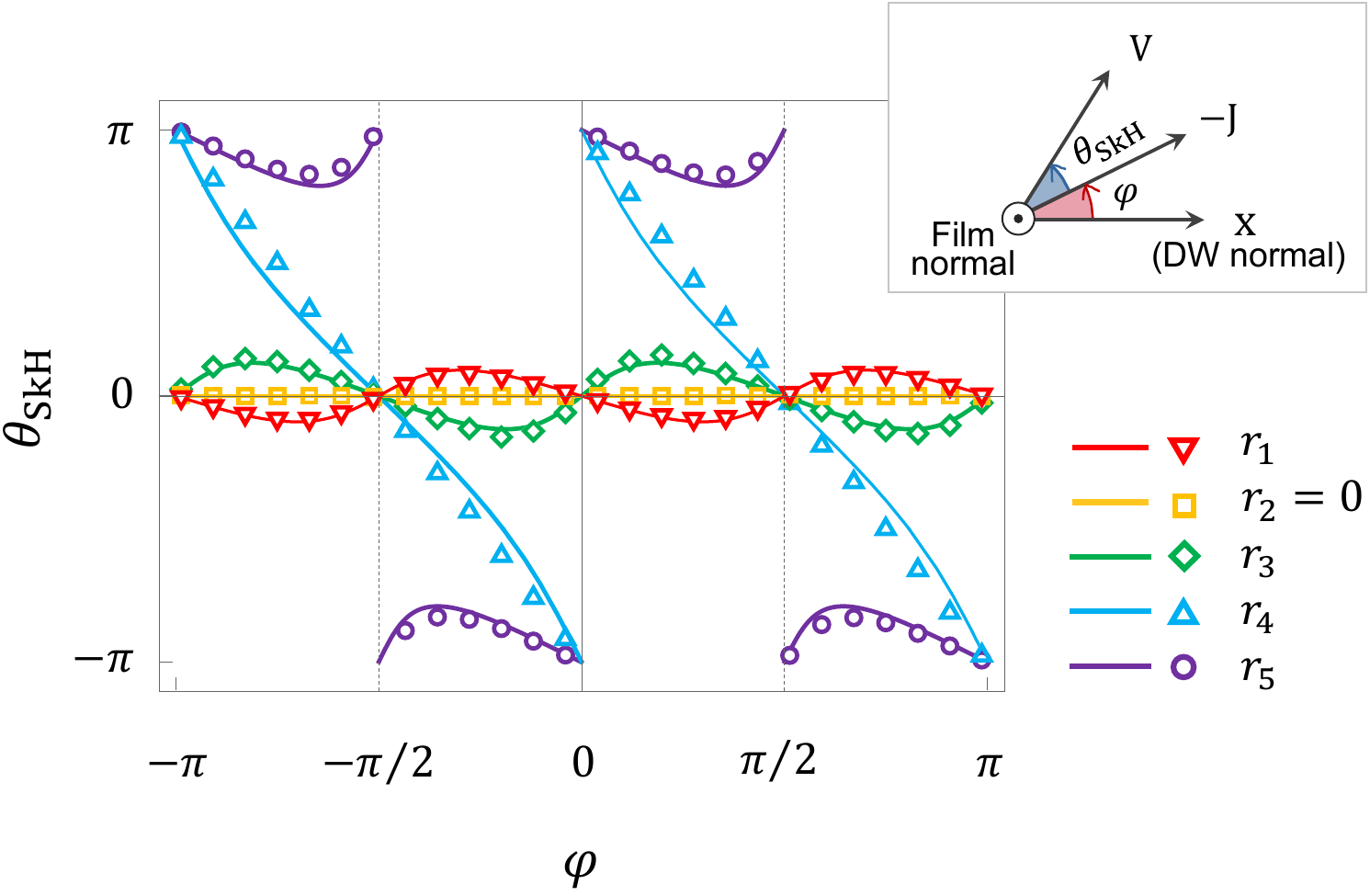}
    \caption{
     Skyrmion Hall angle ($\theta_{\mathrm{SkH}}$) as a function of current direction ($\varphi$) [Eq.~(\ref{Eq:HallAngle})]. The inset shows the diagram of these angles. Unlike skyrmions, DW skyrmions exhibit anisotropic skyrmion Hall effect, which reveals five distinct behaviors depending on the value of $r ( = \vartheta / \beta P) $: Some DW skyrmions experience only acute Hall angles (cases $r_1$, $r_3$), others only obtuse $(r_5)$, some across full range of $2\pi$ ($r_4$), and the others none $(r_2)$. Case $r_1$, the only one with positive $r > 0$, experiences a Hall angle in the opposite direction compared to the other cases $r_3, \, r_4,$ and $r_5 \,( < 0 )$. Hall-angle suppression occurs at nodes ($\varphi = 0 , \, \pm \pi/2 , \, \pm \pi$), and in the case $r_2 ( = 0 ) \, ,$  where SOT is absent. Solid lines represent theoretical values while markers indicate simulation results. See the main text for details on the values of $r$ and other material parameters.
    }
    \label{Fig:hallangle}
\end{figure}

In this section, we examine the atypical skyrmion Hall effect of an AFM DW skyrmion. In order to determine its motion, we reduce the Landau-Lifshitz-Gilbert (LLG) equation~\cite{LLG}, a microscopic field equation, to an equation of motion for the center $\bd R(t) = \left( X(t), Y(t) \right)$, a global two-variable equation, by taking a collective coordinate approach~\cite{cc1,cc2}. We will present key intermediate results, leaving out the detailed steps. Our main theoretical results are the current-induced steady-state velocities [Eq.~\eqref{Eq:steadyV}] and the direction of the DW skyrmion motion [\eqref{Eq:HallAngle}], which are compared against numerical simulations in Figs.~\ref{Fig:skhe} and~\ref{Fig:hallangle}, respectively.

We begin with a set of LLG equations for each sublattice $(i = 1, \,  2)$ and additional spin torque terms~\cite{stt1,stt2,stt3,stt4,sot1,sot2,sot3,sot4,sot5,sot6,sot7} given by a current: $s \dot{\bd m}_i - \alpha s \bd m _i \times \dot{\bd m }_i = \bd m _i \times \frac{\delta U}{\delta \bd m _i} + ( P \bd J \cdot \bd{\nabla} ) \bd m_i - \beta \bd m _i \times ( P \bd J \cdot \bd{\nabla} ) \bd m_i - \eta J \bd m_{i} \times \bd p + \vartheta J \bd m_i \times ( \bd p \times \bd m_i) \, ,$ which translates into the following equation in terms of the main order parameter $\bd n$:
\begin{widetext}
\begin{equation}
\chi ( 2s )^2 \ddot{\bd n} \times \bd n - 2 \alpha s \bd n \times \dot{\bd n} = 4 s \chi \dot{ \bd n} (\bd H \cdot \bd n ) + \bd n \times \frac{\delta U}{\delta \bd n} - 2 \beta \bd n \times ( P \bd J \cdot \bd{\nabla} ) \bd n + 2 \vartheta J \bd n \times ( \bd p \times \bd n ) \, , \label{Eq:neelEOM}
\end{equation}
\end{widetext}
where $s$ is spin density, $\alpha$ the Gilbert damping parameter, $\beta$ nonadiabaticity, $P$ spin polarization, $\bd H = 2 \eta J \bd p$ effective field given by the field-like spin orbit torque (SOT)$, \vartheta$ damping-like SOT parameter, and $\bd p = \hat{\bd z} \times \bd J / | \bd J | \, .$ Here, the external magnetic field is not considered, since it proved to induce no motion as discussed later. By capturing the time dynamics of the order parameter through the dynamics of the collective coordinates, $\bd n ( \bd r, t) = \bd n(\bd r; \bd R(t)) $, Eq.~\eqref{Eq:neelEOM} can be recast into
\begin{equation}
\chi ( 2 s ) ^2 \mathcal{M}\ddot{\bd R} + 2 \alpha s \mathcal{M} \dot{\bd R} = \mathcal{P}\bd J \, , \label{Eq:EOM}
\end{equation}
where the matrix $\mathcal P =  2 \beta P \mathcal M + 2 \vartheta \Theta $ represents the effect of spin torques, which includes a nonadiabatic STT ($\propto \beta P$) and a damping-like SOT ($\propto \vartheta$). This is done by assuming the rigid arrangement of the order parameter $\bd n ( \bd r )$, which is valid in low-energy dynamics~\cite{cc1,cc2}. Here, matrices $\mathcal M$ and $\Theta$ are characteristic values of the spin structure: $\mathcal M _{ij} = \iint \partial_i \bd n \cdot \partial_j \bd n \,\,\, dx \, dy \,$, and $\Theta_{ij} = \iint ( \bd n \times \partial_i \bd n ) \cdot \hat{\bd x}_{j} \times \hat{\bd z}  \,\,\, dx \, dy \, $, evaluating to $\mathcal M = \mathrm{Diag} \left( \frac{2 \Delta y}{\ld} , \,  \frac{16}{3}\Big( \frac{\ld}{\ls} \Big) ^3 + 16 \cdot \frac{\ld}{\ls} \right) \,$, and $\Theta = \mathrm{Diag} \left( 2 \pi ( \Delta y - 4 \ls ) \, , 16 \pi \ld^{\,\,2} / (3 \ls) \right)$, where $\Delta y$ is a length of a system along a DW. Note that $\mathcal P$ lacks the other two channels through which a current drives the motion, which will be discussed later.

In a steady state, the velocity reaches
 \begin{equation}
 \pmat{ V_x \\ V_y } = \frac{ -1 }{\alpha s} \pmat{ \beta P + \vartheta \frac{ \Theta_{11} }{\mathcal{M}_{11}} & 0 \\ 0 & \beta P + \vartheta \frac{\Theta_{22}}{\mathcal{M}_{22}} } \pmat{ J_{x} \\ J_{y} } \, , \label{Eq:steadyV}
 \end{equation}
in response to a given current density $\bd J = ( J_x , \, J_y )$, where the coefficient matrix includes a couple of spin torque parameters: $\beta P$, a non-adiabatic STT and $\vartheta$, a damping-like SOT. This is our first main theoretical result, which implies an atypical skyrmion Hall effect of a DW skyrmion. Before discussing the effect, let us take a look at some features. Several components that one might expect is missing in Eq.~\eqref{Eq:steadyV}: a couple of the other spin torque parameters—$P$ and $\eta$—, and topological quantities. (1)~The adiabatic STT ($\propto P$) vanishes due to the transfer of opposite spins to each sublattice, resulting in cancellation. (2)~The field-like SOT ($\propto \eta$) vanishes due to the integral skyrmion charge of each sublattice. We mention in passing that, if the sublattices had fractional charge, a magnetic field—whether internal or external—could induce a finite net skyrmion charge~\cite{Dasgupta}, then both of the adiabatic STT and the field-like SOT could induce its motion. (3)~The absence of topological quantity suggests that two DW skyrmions with opposite windings follow the same equation of motion and exhibit the same dynamics. Now let us focus on that the coefficient matrix is diagonal but may have different elements. This indicates that DW skyrmions can move at an angle to a current: a skyrmion Hall effect~\cite{skyrmion1,skyrmion2,skyrmion3}. Yet it is atypical in that, the effect comes from asymmetric spin configuration where $\Theta_{11} / \mathcal{M}_{11} \neq \Theta_{22} / \mathcal{M}_{22}$, not from topological properties. An asymmetric system discerns the direction of the spin current $\hat{\bd p} = \hat{\bd z} \times \bd J / | \bd J|$ via SOT. Moreover, it is atypical in that the effect can be suppressed if a current is applied either in the $x$ (DW normal) or in the $y$ (DW parallel) direction. In a sense, the DW normal and the DW parallel are the principal axes of the spin structure of its current-driven motion. I.e., current in $x$- ($y$-) direction, $J_x \,\,(J_y)$, cannot induce the motion in $y$- ($x$-) direction, $V_y \,\, (V_x)$. Specifically, the suppression along a DW ($y$-axis) is noteworthy as it facilitates the use of a DW skyrmion in DW-substructure-based memory~\cite{yang}. Figure~\ref{Fig:skhe} shows the suppression and manifestation of the effect, which is our theoretical and simulation results. The simulation details and material parameters used here will be provided at the end of the section.

Unlike the radially symmetric skyrmions, where no current direction is unique, a DW skyrmion is expected to show a skyrmion Hall effect that varies with current direction. The Hall angle is given by
\begin{widetext}
\begin{equation}
\theta_{ \mathrm{SkH} } ( \varphi ) = \mathrm{atan2} \Bigg[ 1 + r \left( \A \cos ^2 \varphi + \B \sin ^2 \varphi \right) , \, - r \left( \A - \B \right) \sin \varphi \cos \varphi \Bigg] \label{Eq:HallAngle}
\end{equation}
\end{widetext}
as a function of the current direction, which is our second main result. Here, $\varphi$ represents the current direction such that $ - \bd{J} = | \bd J | \cos \varphi \, \hat{\bd x} + | \bd J | \sin \varphi \, \hat{\bd y}$, and the effects of the two spin torques are incorporated through their ratio $r = \vartheta / (\beta P)$. The two-argument $\arctan$ function, $\mathrm{atan2} \, ,$ has a range of $ 2\pi$, enabling it to distinguish the phases of the points $ (x, \, y) $ and $ ( -x, \, -y ) $, in contrast to the original $\arctan(y/x)$ that does not distinguish $(x,y)$ and $(-x, -y)$. To better understand Eq.~\eqref{Eq:HallAngle}, let us refer to Fig.~\ref{Fig:hallangle}. Here, we examined the case where $\beta P > 0$ and $\frac{\Theta_{11}}{\mathcal{M}_{11}} > \frac{\Theta_{22}}{\mathcal{M}_{22}}$, which exemplifies the general trend. Note that five different behaviors are classified according to the value of $r = \vartheta / (\beta P)$, which are separated by the critical values $r = 0, - \mathcal{M}_{11} / \Theta_{11}$, and $- \mathcal{M}_{22} / \Theta_{22}$, satisfying $r_1 > r_2 = 0 > r_3 > - \mathcal{M}_{22}/\Theta_{22} > r_4 > - \mathcal{M}_{11} / \Theta_{11} > r_5 \,.$

To be specific, the Hall angle range that a DW skyrmion experiences differs depending on the value of $r$: acute angles only (in $r_1$ and $r_3$ regime), obtuse angle only ($r_5$), the full range of $2\pi$ ($r_4$), or none ($r_2 = 0$). The acuteness or obtuseness of the angle depends on the sign of longitudinal velocity ($V_{\parallel} \equiv \bd V \cdot \bd J / | \bd J |$ ), whether positive or negative, which is given by $V_{\parallel} = \frac{\beta P J}{\alpha s} + \frac{\vartheta J}{\alpha s} \left( \A \cos ^2 \varphi + \B \sin ^2 \varphi \right) \, .$ The longitudinal velocity can be decomposed into contributions from STT and SOT. Let $V_{\parallel}^{\mathrm{stt}}$ be fixed (positive, given that $\beta P > 0 $). With changing the value of $r$, $V_{\parallel}^{\mathrm{sot}}$ varies and oscillates as current direction $\varphi$ changes. The behavior of $\mathrm{sgn} ( V_{\parallel} )$ is determined by the rivalry between STT and SOT contribution: $V_{\parallel} ^{\mathrm{sot}}$ can either reinforce $V_{\parallel} ^{\mathrm{stt}}$ or compete with it, in which case $V_{\parallel}^{\mathrm{stt}}$ dominates, $V_{\parallel}^{\mathrm{sot}}$ dominates, or they are comparable in strength such that the dominance changes with the oscillation. Each corresponds to the case $r_1 , \,  r_3 , \, r_5 ,$ and $ r_4 ,$ respectively. In addition, the sign of the Hall angle is the same as the sign of the transverse velocity ($V_{\bot} \equiv \bd V \cdot \hat{\bd z} \times \bd J / | \bd J| $), which is given by $V_{\bot} = - \frac{\vartheta J}{\alpha s} \left( \A - \B \right) \sin \varphi \cos \varphi$. Note that the finite $V_{\bot}$, or the Skyrmion Hall effect, is exclusively induced by SOT. Case $r_2$, in which SOT is absent, shows the vanishing Hall angle. Case $r_1$, the only one with the opposite sign ($r_1 > 0$), shows a Hall angle with the opposite sign to that of cases $r_3$, $r_4$, and $r_5 ( < 0 )$. Last, observe the nodes at $\varphi = 0 , \, \pm \pi/2 , \, \pm \pi$ across all cases. The effect is suppressed when the current is applied either along or transverse to a DW, and the principal axes are determined solely by a spin configuration, independent of other material parameters such as $r$. Note that the principal axes correspond to the axis of geometric symmetry (DW normal) and its perpendicular axis (DW parallel).

Lastly, we provide simulation details and material parameters we used to plot Figs.~\ref{Fig:skhe} and \ref{Fig:hallangle}. We used mumax$^3$ micromagnetic simulations ~\cite{mumax3(1),mumax3(2)}. The size of the sample was $250 \mathrm{nm} \times 125 \mathrm{nm} \times 1 \mathrm{nm}$ with $256 \times 128 \times 1$ lattice points (presented in the order of $x , \, y ,$ and $z$ direction) and periodic boundary condition was adopted in the $y$ direction. Material parameters are as follows: $\ms = 580 \mathrm{kA/m} , \, A =  1.5 \times 10^{11} \mathrm{J/m} , \, K = 8 \times 10^5 \mathrm{J/m^3} , \, D = 1.1 \times 10^{-3} \mathrm{J/m^2}$. Here, we set $A_{\mathrm{sim}} = - A$ in order to model Heisenberg AFM thin films in mumax$^3$, which has been proven to be suitable in previous studies, e.g., Ref.~\cite{Asim}. For dynamics, $\alpha = 0.1 , \, P = 0.4\hbar / | 2 e | $ with electron charge $e$. Nonadiabaticity STT parameter was fixed to $\beta = 0.3$ throughout the simulation, while SOT parameter $\vartheta = r \beta P $ varies with different values of $r$. In (d), $r$ takes several values: $r_1 = 8.33 , \, r_2 = 0 , \, r_3 = -1.39 , \, r_4 = -2.50 , \, r_5 = - 8.33$ in the unit of $10^{8} \mathrm{m^{-1}}$ where the critical values are $r = 0, \, - 1.71 , \, -5.88$. If not specified, $r = 8.33 \times 10^8 \mathrm{m^{-1}}$ and current density $J = 1.0 \times 10^{11} \mathrm{A/m^2}$.

\section{Conclusion}
\label{sec:last}
Our study reveals that DW skyrmions can be stabilized in AFMs and exhibit a conditional skyrmion Hall effect during current-driven motion. Besides the topology, the asymmetry of a spin configuration can also give rise to a skyrmion Hall effect, which is well exemplified by the contrasting behavior of an AFM DW skyrmion (the asymmetric) and an AFM skyrmion (the symmetric): only the former exhibits the Hall effect despite of the same topology. We understand this anisotropy-based Hall effect through the concept of principal axes. The motion of a DW skyrmion becomes misaligned with the current unless it is applied along one of these principal axes, leading to the skyrmion Hall effect. Specifically, the current-aligned motion along the specific principal axis, a DW, and their versatility in AFM systems make DW skyrmions a promising candidate for DW-based memory devices. We hope our work on the current-induced dynamics of AFM DW skyrmions triggers further research on the exotic dynamics of composite magnetic solitons and resultant novel functionalities.

\begin{acknowledgments} We thank Soong-Geun Je for the insightful discussions. This work was supported by the Brain Pool Plus Program through the National Research Foundation of Korea funded by the Ministry of Science and ICT (NRF-2020H1D3A2A03099291).
\end{acknowledgments}

\bibliographystyle{apsrev4-2}
\bibliography{bib.bib}

\begin{thebibliography}{32}%
\makeatletter
\providecommand \@ifxundefined [1]{%
 \@ifx{#1\undefined}
}%
\providecommand \@ifnum [1]{%
 \ifnum #1\expandafter \@firstoftwo
 \else \expandafter \@secondoftwo
 \fi
}%
\providecommand \@ifx [1]{%
 \ifx #1\expandafter \@firstoftwo
 \else \expandafter \@secondoftwo
 \fi
}%
\providecommand \natexlab [1]{#1}%
\providecommand \enquote  [1]{``#1''}%
\providecommand \bibnamefont  [1]{#1}%
\providecommand \bibfnamefont [1]{#1}%
\providecommand \citenamefont [1]{#1}%
\providecommand \href@noop [0]{\@secondoftwo}%
\providecommand \href [0]{\begingroup \@sanitize@url \@href}%
\providecommand \@href[1]{\@@startlink{#1}\@@href}%
\providecommand \@@href[1]{\endgroup#1\@@endlink}%
\providecommand \@sanitize@url [0]{\catcode `\\12\catcode `\$12\catcode `\&12\catcode `\#12\catcode `\^12\catcode `\_12\catcode `\%12\relax}%
\providecommand \@@startlink[1]{}%
\providecommand \@@endlink[0]{}%
\providecommand \url  [0]{\begingroup\@sanitize@url \@url }%
\providecommand \@url [1]{\endgroup\@href {#1}{\urlprefix }}%
\providecommand \urlprefix  [0]{URL }%
\providecommand \Eprint [0]{\href }%
\providecommand \doibase [0]{https://doi.org/}%
\providecommand \selectlanguage [0]{\@gobble}%
\providecommand \bibinfo  [0]{\@secondoftwo}%
\providecommand \bibfield  [0]{\@secondoftwo}%
\providecommand \translation [1]{[#1]}%
\providecommand \BibitemOpen [0]{}%
\providecommand \bibitemStop [0]{}%
\providecommand \bibitemNoStop [0]{.\EOS\space}%
\providecommand \EOS [0]{\spacefactor3000\relax}%
\providecommand \BibitemShut  [1]{\csname bibitem#1\endcsname}%
\let\auto@bib@innerbib\@empty
%</preamble>
\bibitem [{\citenamefont {Kosevich}\ \emph {et~al.}(1990)\citenamefont {Kosevich}, \citenamefont {Ivanov},\ and\ \citenamefont {Kovalev}}]{KosevichPR1990}%
  \BibitemOpen
  \bibfield  {author} {\bibinfo {author} {\bibfnamefont {A.}~\bibnamefont {Kosevich}}, \bibinfo {author} {\bibfnamefont {B.}~\bibnamefont {Ivanov}},\ and\ \bibinfo {author} {\bibfnamefont {A.}~\bibnamefont {Kovalev}},\ }\href {https://doi.org/http://dx.doi.org/10.1016/0370-1573(90)90130-T} {\bibfield  {journal} {\bibinfo  {journal} {Phys. Rep.}\ }\textbf {\bibinfo {volume} {194}},\ \bibinfo {pages} {117 } (\bibinfo {year} {1990})}\BibitemShut {NoStop}%
\bibitem [{\citenamefont {Galkina}\ and\ \citenamefont {Ivanov}(2018)}]{GalkinaLTP2018}%
  \BibitemOpen
  \bibfield  {author} {\bibinfo {author} {\bibfnamefont {E.~G.}\ \bibnamefont {Galkina}}\ and\ \bibinfo {author} {\bibfnamefont {B.~A.}\ \bibnamefont {Ivanov}},\ }\href {https://doi.org/10.1063/1.5041427} {\bibfield  {journal} {\bibinfo  {journal} {Low Temp. Phys.}\ }\textbf {\bibinfo {volume} {44}},\ \bibinfo {pages} {618} (\bibinfo {year} {2018})}\BibitemShut {NoStop}%
\bibitem [{\citenamefont {Yoshimura}\ \emph {et~al.}(2016)\citenamefont {Yoshimura}, \citenamefont {Kim}, \citenamefont {Taniguchi}, \citenamefont {Tono}, \citenamefont {Ueda}, \citenamefont {Hiramatsu}, \citenamefont {Moriyama}, \citenamefont {Yamada}, \citenamefont {Nakatani},\ and\ \citenamefont {Ono}}]{Kab}%
  \BibitemOpen
  \bibfield  {author} {\bibinfo {author} {\bibfnamefont {Y.}~\bibnamefont {Yoshimura}}, \bibinfo {author} {\bibfnamefont {K.-J.}\ \bibnamefont {Kim}}, \bibinfo {author} {\bibfnamefont {T.}~\bibnamefont {Taniguchi}}, \bibinfo {author} {\bibfnamefont {T.}~\bibnamefont {Tono}}, \bibinfo {author} {\bibfnamefont {K.}~\bibnamefont {Ueda}}, \bibinfo {author} {\bibfnamefont {R.}~\bibnamefont {Hiramatsu}}, \bibinfo {author} {\bibfnamefont {T.}~\bibnamefont {Moriyama}}, \bibinfo {author} {\bibfnamefont {K.}~\bibnamefont {Yamada}}, \bibinfo {author} {\bibfnamefont {Y.}~\bibnamefont {Nakatani}},\ and\ \bibinfo {author} {\bibfnamefont {T.}~\bibnamefont {Ono}},\ }\href {https://www.nature.com/articles/nphys3535} {\bibfield  {journal} {\bibinfo  {journal} {Nat. Phys.}\ }\textbf {\bibinfo {volume} {12}},\ \bibinfo {pages} {157} (\bibinfo {year} {2016})}\BibitemShut {NoStop}%
\bibitem [{\citenamefont {Parkin}\ and\ \citenamefont {Yang}(2015)}]{racetrack}%
  \BibitemOpen
  \bibfield  {author} {\bibinfo {author} {\bibfnamefont {S.}~\bibnamefont {Parkin}}\ and\ \bibinfo {author} {\bibfnamefont {S.-H.}\ \bibnamefont {Yang}},\ }\href@noop {} {\bibfield  {journal} {\bibinfo  {journal} {Nat. Nanotechnol.}\ }\textbf {\bibinfo {volume} {10}},\ \bibinfo {pages} {195} (\bibinfo {year} {2015})}\BibitemShut {NoStop}%
\bibitem [{\citenamefont {Yang}\ \emph {et~al.}(2021)\citenamefont {Yang}, \citenamefont {Moon}, \citenamefont {Park}, \citenamefont {Lee}, \citenamefont {Kang}, \citenamefont {Shin}, \citenamefont {Kim},\ and\ \citenamefont {Kim}}]{yang}%
  \BibitemOpen
  \bibfield  {author} {\bibinfo {author} {\bibfnamefont {J.}~\bibnamefont {Yang}}, \bibinfo {author} {\bibfnamefont {K.-W.}\ \bibnamefont {Moon}}, \bibinfo {author} {\bibfnamefont {A.~M.~G.}\ \bibnamefont {Park}}, \bibinfo {author} {\bibfnamefont {S.}~\bibnamefont {Lee}}, \bibinfo {author} {\bibfnamefont {D.~H.}\ \bibnamefont {Kang}}, \bibinfo {author} {\bibfnamefont {M.}~\bibnamefont {Shin}}, \bibinfo {author} {\bibfnamefont {S.}~\bibnamefont {Kim}},\ and\ \bibinfo {author} {\bibfnamefont {K.-J.}\ \bibnamefont {Kim}},\ }\href {https://iopscience.iop.org/article/10.35848/1882-0786/ac2242/meta} {\bibfield  {journal} {\bibinfo  {journal} {Appl. Phys. Express}\ }\textbf {\bibinfo {volume} {14}},\ \bibinfo {pages} {103002} (\bibinfo {year} {2021})}\BibitemShut {NoStop}%
\bibitem [{\citenamefont {Cheng}\ \emph {et~al.}(2019)\citenamefont {Cheng}, \citenamefont {Li}, \citenamefont {Sapkota}, \citenamefont {Rai}, \citenamefont {Pokhrel}, \citenamefont {Mewes}, \citenamefont {Mewes}, \citenamefont {Xiao}, \citenamefont {De~Graef},\ and\ \citenamefont {Sokalski}}]{Cheng}%
  \BibitemOpen
  \bibfield  {author} {\bibinfo {author} {\bibfnamefont {R.}~\bibnamefont {Cheng}}, \bibinfo {author} {\bibfnamefont {M.}~\bibnamefont {Li}}, \bibinfo {author} {\bibfnamefont {A.}~\bibnamefont {Sapkota}}, \bibinfo {author} {\bibfnamefont {A.}~\bibnamefont {Rai}}, \bibinfo {author} {\bibfnamefont {A.}~\bibnamefont {Pokhrel}}, \bibinfo {author} {\bibfnamefont {T.}~\bibnamefont {Mewes}}, \bibinfo {author} {\bibfnamefont {C.}~\bibnamefont {Mewes}}, \bibinfo {author} {\bibfnamefont {D.}~\bibnamefont {Xiao}}, \bibinfo {author} {\bibfnamefont {M.}~\bibnamefont {De~Graef}},\ and\ \bibinfo {author} {\bibfnamefont {V.}~\bibnamefont {Sokalski}},\ }\href {https://journals.aps.org/prb/abstract/10.1103/PhysRevB.99.184412} {\bibfield  {journal} {\bibinfo  {journal} {Phys. Rev. B}\ }\textbf {\bibinfo {volume} {99}},\ \bibinfo {pages} {184412} (\bibinfo {year} {2019})}\BibitemShut {NoStop}%
\bibitem [{\citenamefont {Ross}\ and\ \citenamefont {Nitta}(2023)}]{Nitta}%
  \BibitemOpen
  \bibfield  {author} {\bibinfo {author} {\bibfnamefont {C.}~\bibnamefont {Ross}}\ and\ \bibinfo {author} {\bibfnamefont {M.}~\bibnamefont {Nitta}},\ }\href {https://link.aps.org/pdf/10.1103/PhysRevB.107.024422} {\bibfield  {journal} {\bibinfo  {journal} {Phys. Rev. B}\ }\textbf {\bibinfo {volume} {107}},\ \bibinfo {pages} {024422} (\bibinfo {year} {2023})}\BibitemShut {NoStop}%
\bibitem [{\citenamefont {Han}\ \emph {et~al.}(2024)\citenamefont {Han}, \citenamefont {Kim}, \citenamefont {Kim},\ and\ \citenamefont {Je}}]{Je}%
  \BibitemOpen
  \bibfield  {author} {\bibinfo {author} {\bibfnamefont {S.~U.}\ \bibnamefont {Han}}, \bibinfo {author} {\bibfnamefont {W.}~\bibnamefont {Kim}}, \bibinfo {author} {\bibfnamefont {S.~K.}\ \bibnamefont {Kim}},\ and\ \bibinfo {author} {\bibfnamefont {S.-G.}\ \bibnamefont {Je}},\ }\href {https://journals.aps.org/prb/abstract/10.1103/PhysRevB.109.014404} {\bibfield  {journal} {\bibinfo  {journal} {Phys. Rev. B}\ }\textbf {\bibinfo {volume} {109}},\ \bibinfo {pages} {014404} (\bibinfo {year} {2024})}\BibitemShut {NoStop}%
\bibitem [{\citenamefont {Landau}\ and\ \citenamefont {Lifshitz}(2013)}]{LLG}%
  \BibitemOpen
  \bibfield  {author} {\bibinfo {author} {\bibfnamefont {L.~D.}\ \bibnamefont {Landau}}\ and\ \bibinfo {author} {\bibfnamefont {E.~M.}\ \bibnamefont {Lifshitz}},\ }\href@noop {} {\emph {\bibinfo {title} {Quantum mechanics: non-relativistic theory}}},\ Vol.~\bibinfo {volume} {3}\ (\bibinfo  {publisher} {Elsevier},\ \bibinfo {year} {2013})\BibitemShut {NoStop}%
\bibitem [{\citenamefont {De~Clercq}\ \emph {et~al.}(2016)\citenamefont {De~Clercq}, \citenamefont {Vansteenkiste}, \citenamefont {Abes}, \citenamefont {Temst},\ and\ \citenamefont {Van~Waeyenberge}}]{mumax3(1)}%
  \BibitemOpen
  \bibfield  {author} {\bibinfo {author} {\bibfnamefont {J.}~\bibnamefont {De~Clercq}}, \bibinfo {author} {\bibfnamefont {A.}~\bibnamefont {Vansteenkiste}}, \bibinfo {author} {\bibfnamefont {M.}~\bibnamefont {Abes}}, \bibinfo {author} {\bibfnamefont {K.}~\bibnamefont {Temst}},\ and\ \bibinfo {author} {\bibfnamefont {B.}~\bibnamefont {Van~Waeyenberge}},\ }\href {https://iopscience.iop.org/article/10.1088/0022-3727/49/43/435001/meta} {\bibfield  {journal} {\bibinfo  {journal} {J. Phys. D:Appl. Phys.}\ }\textbf {\bibinfo {volume} {49}},\ \bibinfo {pages} {435001} (\bibinfo {year} {2016})}\BibitemShut {NoStop}%
\bibitem [{\citenamefont {Mulkers}\ \emph {et~al.}(2017)\citenamefont {Mulkers}, \citenamefont {Van~Waeyenberge},\ and\ \citenamefont {Milo{\v{s}}evi{\'c}}}]{mumax3(2)}%
  \BibitemOpen
  \bibfield  {author} {\bibinfo {author} {\bibfnamefont {J.}~\bibnamefont {Mulkers}}, \bibinfo {author} {\bibfnamefont {B.}~\bibnamefont {Van~Waeyenberge}},\ and\ \bibinfo {author} {\bibfnamefont {M.~V.}\ \bibnamefont {Milo{\v{s}}evi{\'c}}},\ }\href {https://journals.aps.org/prb/abstract/10.1103/PhysRevB.95.144401} {\bibfield  {journal} {\bibinfo  {journal} {Phys. Rev. B}\ }\textbf {\bibinfo {volume} {95}},\ \bibinfo {pages} {144401} (\bibinfo {year} {2017})}\BibitemShut {NoStop}%
\bibitem [{\citenamefont {Berger}(1996)}]{stt1}%
  \BibitemOpen
  \bibfield  {author} {\bibinfo {author} {\bibfnamefont {L.}~\bibnamefont {Berger}},\ }\href {https://link.aps.org/pdf/10.1103/PhysRevB.54.9353} {\bibfield  {journal} {\bibinfo  {journal} {Phys. Rev. B}\ }\textbf {\bibinfo {volume} {54}},\ \bibinfo {pages} {9353} (\bibinfo {year} {1996})}\BibitemShut {NoStop}%
\bibitem [{\citenamefont {Slonczewski}(1996)}]{stt2}%
  \BibitemOpen
  \bibfield  {author} {\bibinfo {author} {\bibfnamefont {J.~C.}\ \bibnamefont {Slonczewski}},\ }\href {https://www.sciencedirect.com/science/article/pii/0304885396000625} {\bibfield  {journal} {\bibinfo  {journal} {J. Magn. Magn. Mater.}\ }\textbf {\bibinfo {volume} {159}},\ \bibinfo {pages} {L1} (\bibinfo {year} {1996})}\BibitemShut {NoStop}%
\bibitem [{\citenamefont {Brataas}\ \emph {et~al.}(2012)\citenamefont {Brataas}, \citenamefont {Kent},\ and\ \citenamefont {Ohno}}]{stt3}%
  \BibitemOpen
  \bibfield  {author} {\bibinfo {author} {\bibfnamefont {A.}~\bibnamefont {Brataas}}, \bibinfo {author} {\bibfnamefont {A.~D.}\ \bibnamefont {Kent}},\ and\ \bibinfo {author} {\bibfnamefont {H.}~\bibnamefont {Ohno}},\ }\href {https://www.nature.com/articles/nmat3311} {\bibfield  {journal} {\bibinfo  {journal} {Nat. Mater.}\ }\textbf {\bibinfo {volume} {11}},\ \bibinfo {pages} {372} (\bibinfo {year} {2012})}\BibitemShut {NoStop}%
\bibitem [{\citenamefont {Kawahara}\ \emph {et~al.}(2012)\citenamefont {Kawahara}, \citenamefont {Ito}, \citenamefont {Takemura},\ and\ \citenamefont {Ohno}}]{stt4}%
  \BibitemOpen
  \bibfield  {author} {\bibinfo {author} {\bibfnamefont {T.}~\bibnamefont {Kawahara}}, \bibinfo {author} {\bibfnamefont {K.}~\bibnamefont {Ito}}, \bibinfo {author} {\bibfnamefont {R.}~\bibnamefont {Takemura}},\ and\ \bibinfo {author} {\bibfnamefont {H.}~\bibnamefont {Ohno}},\ }\href {https://www.sciencedirect.com/science/article/pii/S002627141100446X} {\bibfield  {journal} {\bibinfo  {journal} {Microelectron. Reliab.}\ }\textbf {\bibinfo {volume} {52}},\ \bibinfo {pages} {613} (\bibinfo {year} {2012})}\BibitemShut {NoStop}%
\bibitem [{\citenamefont {Manchon}\ \emph {et~al.}(2019)\citenamefont {Manchon}, \citenamefont {{\v{Z}}elezn{\`y}}, \citenamefont {Miron}, \citenamefont {Jungwirth}, \citenamefont {Sinova}, \citenamefont {Thiaville}, \citenamefont {Garello},\ and\ \citenamefont {Gambardella}}]{sot1}%
  \BibitemOpen
  \bibfield  {author} {\bibinfo {author} {\bibfnamefont {A.}~\bibnamefont {Manchon}}, \bibinfo {author} {\bibfnamefont {J.}~\bibnamefont {{\v{Z}}elezn{\`y}}}, \bibinfo {author} {\bibfnamefont {I.~M.}\ \bibnamefont {Miron}}, \bibinfo {author} {\bibfnamefont {T.}~\bibnamefont {Jungwirth}}, \bibinfo {author} {\bibfnamefont {J.}~\bibnamefont {Sinova}}, \bibinfo {author} {\bibfnamefont {A.}~\bibnamefont {Thiaville}}, \bibinfo {author} {\bibfnamefont {K.}~\bibnamefont {Garello}},\ and\ \bibinfo {author} {\bibfnamefont {P.}~\bibnamefont {Gambardella}},\ }\href {https://journals.aps.org/rmp/abstract/10.1103/RevModPhys.91.035004} {\bibfield  {journal} {\bibinfo  {journal} {Rev. Mod. Phys.}\ }\textbf {\bibinfo {volume} {91}},\ \bibinfo {pages} {035004} (\bibinfo {year} {2019})}\BibitemShut {NoStop}%
\bibitem [{\citenamefont {Altland}\ and\ \citenamefont {Simons}(2010)}]{sot2}%
  \BibitemOpen
  \bibfield  {author} {\bibinfo {author} {\bibfnamefont {A.}~\bibnamefont {Altland}}\ and\ \bibinfo {author} {\bibfnamefont {B.~D.}\ \bibnamefont {Simons}},\ }\href@noop {} {\emph {\bibinfo {title} {Condensed matter field theory}}}\ (\bibinfo  {publisher} {Cambridge university press},\ \bibinfo {year} {2010})\BibitemShut {NoStop}%
\bibitem [{\citenamefont {Tserkovnyak}\ and\ \citenamefont {Bender}(2014)}]{sot3}%
  \BibitemOpen
  \bibfield  {author} {\bibinfo {author} {\bibfnamefont {Y.}~\bibnamefont {Tserkovnyak}}\ and\ \bibinfo {author} {\bibfnamefont {S.~A.}\ \bibnamefont {Bender}},\ }\href {https://journals.aps.org/prb/abstract/10.1103/PhysRevB.90.014428} {\bibfield  {journal} {\bibinfo  {journal} {Phys. Rev. B}\ }\textbf {\bibinfo {volume} {90}},\ \bibinfo {pages} {014428} (\bibinfo {year} {2014})}\BibitemShut {NoStop}%
\bibitem [{\citenamefont {Edelstein}(1990)}]{sot4}%
  \BibitemOpen
  \bibfield  {author} {\bibinfo {author} {\bibfnamefont {V.~M.}\ \bibnamefont {Edelstein}},\ }\href {https://www.sciencedirect.com/science/article/pii/003810989090963C} {\bibfield  {journal} {\bibinfo  {journal} {Solid State Commun.}\ }\textbf {\bibinfo {volume} {73}},\ \bibinfo {pages} {233} (\bibinfo {year} {1990})}\BibitemShut {NoStop}%
\bibitem [{\citenamefont {Manchon}\ \emph {et~al.}(2015)\citenamefont {Manchon}, \citenamefont {Koo}, \citenamefont {Nitta}, \citenamefont {Frolov},\ and\ \citenamefont {Duine}}]{sot5}%
  \BibitemOpen
  \bibfield  {author} {\bibinfo {author} {\bibfnamefont {A.}~\bibnamefont {Manchon}}, \bibinfo {author} {\bibfnamefont {H.~C.}\ \bibnamefont {Koo}}, \bibinfo {author} {\bibfnamefont {J.}~\bibnamefont {Nitta}}, \bibinfo {author} {\bibfnamefont {S.~M.}\ \bibnamefont {Frolov}},\ and\ \bibinfo {author} {\bibfnamefont {R.~A.}\ \bibnamefont {Duine}},\ }\href {https://www.nature.com/articles/nmat4360} {\bibfield  {journal} {\bibinfo  {journal} {Nat. Mater.}\ }\textbf {\bibinfo {volume} {14}},\ \bibinfo {pages} {871} (\bibinfo {year} {2015})}\BibitemShut {NoStop}%
\bibitem [{\citenamefont {Dyakonov}(1971)}]{sot6}%
  \BibitemOpen
  \bibfield  {author} {\bibinfo {author} {\bibfnamefont {M.~I.}\ \bibnamefont {Dyakonov}},\ }\href@noop {} {\bibfield  {journal} {\bibinfo  {journal} {JETP Lett. USSR}\ }\textbf {\bibinfo {volume} {13}},\ \bibinfo {pages} {467} (\bibinfo {year} {1971})}\BibitemShut {NoStop}%
\bibitem [{\citenamefont {Sinova}\ \emph {et~al.}(2015)\citenamefont {Sinova}, \citenamefont {Valenzuela}, \citenamefont {Wunderlich}, \citenamefont {Back},\ and\ \citenamefont {Jungwirth}}]{sot7}%
  \BibitemOpen
  \bibfield  {author} {\bibinfo {author} {\bibfnamefont {J.}~\bibnamefont {Sinova}}, \bibinfo {author} {\bibfnamefont {S.~O.}\ \bibnamefont {Valenzuela}}, \bibinfo {author} {\bibfnamefont {J.}~\bibnamefont {Wunderlich}}, \bibinfo {author} {\bibfnamefont {C.}~\bibnamefont {Back}},\ and\ \bibinfo {author} {\bibfnamefont {T.}~\bibnamefont {Jungwirth}},\ }\href {https://journals.aps.org/rmp/abstract/10.1103/RevModPhys.87.1213} {\bibfield  {journal} {\bibinfo  {journal} {Rev. Mod. Phys.}\ }\textbf {\bibinfo {volume} {87}},\ \bibinfo {pages} {1213} (\bibinfo {year} {2015})}\BibitemShut {NoStop}%
\bibitem [{\citenamefont {Dasgupta}\ \emph {et~al.}(2017)\citenamefont {Dasgupta}, \citenamefont {Kim},\ and\ \citenamefont {Tchernyshyov}}]{Dasgupta}%
  \BibitemOpen
  \bibfield  {author} {\bibinfo {author} {\bibfnamefont {S.}~\bibnamefont {Dasgupta}}, \bibinfo {author} {\bibfnamefont {S.~K.}\ \bibnamefont {Kim}},\ and\ \bibinfo {author} {\bibfnamefont {O.}~\bibnamefont {Tchernyshyov}},\ }\href {https://link.aps.org/pdf/10.1103/PhysRevB.95.220407} {\bibfield  {journal} {\bibinfo  {journal} {Phys. Rev. B}\ }\textbf {\bibinfo {volume} {95}},\ \bibinfo {pages} {220407} (\bibinfo {year} {2017})}\BibitemShut {NoStop}%
\bibitem [{\citenamefont {Kuchkin}\ \emph {et~al.}(2020)\citenamefont {Kuchkin}, \citenamefont {Barton-Singer}, \citenamefont {Rybakov}, \citenamefont {Bl{\"u}gel}, \citenamefont {Schroers},\ and\ \citenamefont {Kiselev}}]{kink}%
  \BibitemOpen
  \bibfield  {author} {\bibinfo {author} {\bibfnamefont {V.~M.}\ \bibnamefont {Kuchkin}}, \bibinfo {author} {\bibfnamefont {B.}~\bibnamefont {Barton-Singer}}, \bibinfo {author} {\bibfnamefont {F.~N.}\ \bibnamefont {Rybakov}}, \bibinfo {author} {\bibfnamefont {S.}~\bibnamefont {Bl{\"u}gel}}, \bibinfo {author} {\bibfnamefont {B.~J.}\ \bibnamefont {Schroers}},\ and\ \bibinfo {author} {\bibfnamefont {N.~S.}\ \bibnamefont {Kiselev}},\ }\href {https://journals.aps.org/prb/abstract/10.1103/PhysRevB.102.144422} {\bibfield  {journal} {\bibinfo  {journal} {Phys. Rev. B}\ }\textbf {\bibinfo {volume} {102}},\ \bibinfo {pages} {144422} (\bibinfo {year} {2020})}\BibitemShut {NoStop}%
\bibitem [{\citenamefont {Slonczewski}(1972)}]{Slonczewski1}%
  \BibitemOpen
  \bibfield  {author} {\bibinfo {author} {\bibfnamefont {J.}~\bibnamefont {Slonczewski}},\ }\href {https://pubs.aip.org/aip/acp/article/5/1/170/577182/DYNAMICS-OF-MAGNETIC-DOMAIN-WALLS} {\bibfield  {journal} {\bibinfo  {journal} {Magn. Magn. Mater.}\ }\textbf {\bibinfo {volume} {5}},\ \bibinfo {pages} {170} (\bibinfo {year} {1972})}\BibitemShut {NoStop}%
\bibitem [{\citenamefont {Camley}\ and\ \citenamefont {Livesey}(2023)}]{Slonczewski2}%
  \BibitemOpen
  \bibfield  {author} {\bibinfo {author} {\bibfnamefont {R.~E.}\ \bibnamefont {Camley}}\ and\ \bibinfo {author} {\bibfnamefont {K.~L.}\ \bibnamefont {Livesey}},\ }\href {https://www.sciencedirect.com/science/article/pii/S0167572923000201} {\bibfield  {journal} {\bibinfo  {journal} {Surf. Sci. Rep.}\ }\textbf {\bibinfo {volume} {78}},\ \bibinfo {pages} {100605} (\bibinfo {year} {2023})}\BibitemShut {NoStop}%
\bibitem [{\citenamefont {Thiele}(1973)}]{cc1}%
  \BibitemOpen
  \bibfield  {author} {\bibinfo {author} {\bibfnamefont {A.~A.}\ \bibnamefont {Thiele}},\ }\href {https://doi.org/10.1103/PhysRevLett.30.230} {\bibfield  {journal} {\bibinfo  {journal} {Phys. Rev. Lett.}\ }\textbf {\bibinfo {volume} {30}},\ \bibinfo {pages} {230} (\bibinfo {year} {1973})}\BibitemShut {NoStop}%
\bibitem [{\citenamefont {Tretiakov}\ \emph {et~al.}(2008)\citenamefont {Tretiakov}, \citenamefont {Clarke}, \citenamefont {Chern}, \citenamefont {Bazaliy},\ and\ \citenamefont {Tchernyshyov}}]{cc2}%
  \BibitemOpen
  \bibfield  {author} {\bibinfo {author} {\bibfnamefont {O.~A.}\ \bibnamefont {Tretiakov}}, \bibinfo {author} {\bibfnamefont {D.}~\bibnamefont {Clarke}}, \bibinfo {author} {\bibfnamefont {G.-W.}\ \bibnamefont {Chern}}, \bibinfo {author} {\bibfnamefont {Y.~B.}\ \bibnamefont {Bazaliy}},\ and\ \bibinfo {author} {\bibfnamefont {O.}~\bibnamefont {Tchernyshyov}},\ }\href {https://doi.org/10.1103/PhysRevLett.100.127204} {\bibfield  {journal} {\bibinfo  {journal} {Phys. Rev. Lett.}\ }\textbf {\bibinfo {volume} {100}},\ \bibinfo {pages} {127204} (\bibinfo {year} {2008})}\BibitemShut {NoStop}%
\bibitem [{\citenamefont {Nagaosa}\ and\ \citenamefont {Tokura}(2013)}]{skyrmion1}%
  \BibitemOpen
  \bibfield  {author} {\bibinfo {author} {\bibfnamefont {N.}~\bibnamefont {Nagaosa}}\ and\ \bibinfo {author} {\bibfnamefont {Y.}~\bibnamefont {Tokura}},\ }\href {https://www.nature.com/articles/nnano.2013.243} {\bibfield  {journal} {\bibinfo  {journal} {Nat. Nanotechnol.}\ }\textbf {\bibinfo {volume} {8}},\ \bibinfo {pages} {899} (\bibinfo {year} {2013})}\BibitemShut {NoStop}%
\bibitem [{\citenamefont {Everschor-Sitte}\ and\ \citenamefont {Sitte}(2014)}]{skyrmion2}%
  \BibitemOpen
  \bibfield  {author} {\bibinfo {author} {\bibfnamefont {K.}~\bibnamefont {Everschor-Sitte}}\ and\ \bibinfo {author} {\bibfnamefont {M.}~\bibnamefont {Sitte}},\ }\href {https://pubs.aip.org/aip/jap/article/115/17/172602/905394} {\bibfield  {journal} {\bibinfo  {journal} {J. Appl. Phys.}\ }\textbf {\bibinfo {volume} {115}} (\bibinfo {year} {2014})}\BibitemShut {NoStop}%
\bibitem [{\citenamefont {Jiang}\ \emph {et~al.}(2017)\citenamefont {Jiang}, \citenamefont {Zhang}, \citenamefont {Yu}, \citenamefont {Zhang}, \citenamefont {Wang}, \citenamefont {Benjamin~Jungfleisch}, \citenamefont {Pearson}, \citenamefont {Cheng}, \citenamefont {Heinonen}, \citenamefont {Wang} \emph {et~al.}}]{skyrmion3}%
  \BibitemOpen
  \bibfield  {author} {\bibinfo {author} {\bibfnamefont {W.}~\bibnamefont {Jiang}}, \bibinfo {author} {\bibfnamefont {X.}~\bibnamefont {Zhang}}, \bibinfo {author} {\bibfnamefont {G.}~\bibnamefont {Yu}}, \bibinfo {author} {\bibfnamefont {W.}~\bibnamefont {Zhang}}, \bibinfo {author} {\bibfnamefont {X.}~\bibnamefont {Wang}}, \bibinfo {author} {\bibfnamefont {M.}~\bibnamefont {Benjamin~Jungfleisch}}, \bibinfo {author} {\bibfnamefont {J.~E.}\ \bibnamefont {Pearson}}, \bibinfo {author} {\bibfnamefont {X.}~\bibnamefont {Cheng}}, \bibinfo {author} {\bibfnamefont {O.}~\bibnamefont {Heinonen}}, \bibinfo {author} {\bibfnamefont {K.~L.}\ \bibnamefont {Wang}}, \emph {et~al.},\ }\href {https://www.nature.com/articles/nphys3883} {\bibfield  {journal} {\bibinfo  {journal} {Nat. Phys.}\ }\textbf {\bibinfo {volume} {13}},\ \bibinfo {pages} {162} (\bibinfo {year} {2017})}\BibitemShut {NoStop}%
\bibitem [{\citenamefont {Li}\ \emph {et~al.}(2020)\citenamefont {Li}, \citenamefont {Shen}, \citenamefont {Bai}, \citenamefont {Wang}, \citenamefont {Zhang}, \citenamefont {Xia}, \citenamefont {Ezawa}, \citenamefont {Tretiakov}, \citenamefont {Xu}, \citenamefont {Mruczkiewicz} \emph {et~al.}}]{Asim}%
  \BibitemOpen
  \bibfield  {author} {\bibinfo {author} {\bibfnamefont {X.}~\bibnamefont {Li}}, \bibinfo {author} {\bibfnamefont {L.}~\bibnamefont {Shen}}, \bibinfo {author} {\bibfnamefont {Y.}~\bibnamefont {Bai}}, \bibinfo {author} {\bibfnamefont {J.}~\bibnamefont {Wang}}, \bibinfo {author} {\bibfnamefont {X.}~\bibnamefont {Zhang}}, \bibinfo {author} {\bibfnamefont {J.}~\bibnamefont {Xia}}, \bibinfo {author} {\bibfnamefont {M.}~\bibnamefont {Ezawa}}, \bibinfo {author} {\bibfnamefont {O.~A.}\ \bibnamefont {Tretiakov}}, \bibinfo {author} {\bibfnamefont {X.}~\bibnamefont {Xu}}, \bibinfo {author} {\bibfnamefont {M.}~\bibnamefont {Mruczkiewicz}}, \emph {et~al.},\ }\href {https://www.nature.com/articles/s41524-020-00435-y} {\bibfield  {journal} {\bibinfo  {journal} {npj Comput. Mater.}\ }\textbf {\bibinfo {volume} {6}},\ \bibinfo {pages} {169} (\bibinfo {year} {2020})}\BibitemShut {NoStop}%
\end{thebibliography}%

\end{document}